\documentclass[twocolumn,prl,superscriptaddress,showpacs,byrevtex]{revtex4}
\usepackage{epsf,graphicx,amssymb}

\begin{document}
\title{Observation of intermittency in wave turbulence}
\author{E. Falcon}
\email[Corresponding author: ]{eric.falcon@lps.ens.fr}
\altaffiliation[Permanent address: ]{Mati\`ere et Syst\`emes Complexes, Universit\'e Paris 7, CNRS UMR 7057 -- 75 013 Paris, France}
\author{S. Fauve}
\affiliation{Laboratoire de Physique Statistique, \'Ecole Normale Sup\'erieure, UMR 8550, 24, rue Lhomond, 75 005 Paris, France}
\author{C. Laroche}
\affiliation{Laboratoire de Physique, \'Ecole Normale Sup\'erieure de Lyon, UMR 5672, 46, all\'ee d'Italie, 69 007 Lyon, France}

\date{\today}

\begin{abstract}  
We report the observation of intermittency in gravity-capillary wave turbulence on the surface of mercury. We measure the temporal fluctuations of surface wave amplitude at a given location. We show that the shape of the probability density function of the local slope increments of the surface waves strongly changes across the time scales. The related structure functions and the flatness are found to be power laws of the time scale on more than one decade. The exponents of these power laws increase nonlinearly with the order of the structure function. All these observations show the intermittent nature of the increments of the local slope in wave turbulence. We discuss the possible origin of this intermittency.

\end{abstract}

\maketitle
 
One of the most striking feature of turbulence is the occurrence of bursts of intense motion within more quiescent fluid flow. This generates an intermittent behavior \cite{Batchelor49,K62}. One of the quantitative characterization of intermittency is given by the probability density function (PDF) of the velocity increments between two points separated by a distance $r$. Starting from a roughly Gaussian PDF at integral scale, the PDFs undergo a continuous deformation when $r$ is decreased within the inertial range and develop more and more stretched exponential tails \cite{Frisch95}. Deviation from the Gaussian shape can be quantified by the flatness of the PDF. The origin of non-Gaussian statistics in three dimensional hydrodynamic turbulence has been ascribed to the formation of strong vortices since the early work of Batchelor and Townsend \cite{Batchelor49}. However, the physical mechanism of intermittency is still an open question that motivates a lot of studies in three dimensional turbulence \cite{Li05}. Intermittency has been also observed in a lot of problems involving transport by a turbulent flow for which the analytical description of the anomalous scaling laws can be obtained \cite{Falkovitch01}. 

It has been known since the work of Zakharov and collaborators that weakly interacting nonlinear waves can also display Kolmogorov type spectra related to an energy flux cascading from large to small scales \cite{Zakharov67Grav,Zakharov67Cap}. These spectra have been analytically computed using perturbation techniques, but can also be obtained by dimensional analysis using Kolmogorov-type arguments \cite{Connaughton03}. More recently, it has been proposed that intermittency corrections should be also taken into account in wave turbulence \cite{Biven01} and may be connected to singularities or coherent structures \cite{Connaughton03,Choi05} such as wave breaking \cite{Yokoyama04} or whitecaps \cite{Connaughton03} in the case of surface waves. However, intermittency in wave turbulence is often related to non Gaussian statistics of low wave number Fourier amplitudes \cite{Choi05}, thus it is not obviously related to small scale intermittency of hydrodynamic turbulence. Surprisingly, there exist only a small number of experimental studies on wave turbulence \cite{Toba73,Wright96,Lommer02,Brazhnikov02,Onorato04} compared to hydrodynamic turbulence, and to the best of our knowledge, no experimental observation of intermittency has been reported in wave turbulence. 

In this letter, we report the observation of an intermittent behavior for gravity-capillary waves on the surface of a layer of mercury. We show that we need to compute the second-order differences of the surface wave amplitude in order to display intermittency. We observe that the shape of their probability density function changes strongly across the time scales (from a Gaussian at large scales to a stretched exponential shape at short scales). This short-scale intermittency is confirmed by computing the structure functions for various time scales. The structure functions of order $p$ (from 1 to 6) and the flatness are found to be power laws of the time scale on more than one decade. The exponents of the power laws of the structure functions are found to depend nonlinearly on $p$. All these observations show the intermittent nature of the local slope increments of the turbulent surface waves. 

The experimental setup has been already described elsewhere \cite{Falcon06}. It consists of a square vessel, $20 \times 20$ cm$^2$, filled with mercury up to a height of 18 mm. Mercury is chosen because of its low kinematic viscosity (one order of magnitude smaller than that of water), thus reducing wave dissipation. Note however that similar qualitative results to the ones reported here are found when changing mercury by water. Surface waves are generated by the horizontal motion of one rectangular ($13\times 3.5$ cm$^2$) plunging Plexiglas wave maker driven by an electromagnetic vibration exciter. The wave maker is driven with random noise excitation, supplied by a function generator, and selected in a frequency range 0 -  6 Hz by a low-pass filter. The $rms$ value of the velocity fluctuations of the wave maker is proportional to the driving voltage $U_{rms}$ applied to the vibration exciter. Surface waves are generated 3 cm inward from one vessel wall. The local vertical displacement of the fluid  is measured, 7 cm away from the wave maker, by a capacitive sensor. The sensor allows wave height measurements from 10 $\mu$m up to 2 cm with a 20 mm/V sensitivity and a 0.1 ms response time.

\begin{figure}[h]
\centerline{
\epsfysize=65mm
\epsffile{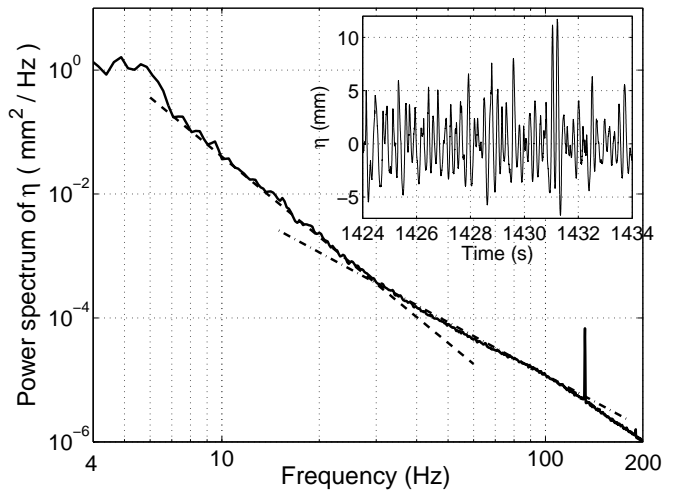} 
}
\caption{Power spectrum of surface wave height, $\eta(t)$. Dashed lines have slopes -4.3 and -2.8. Inset: Typical recording of $\eta(t)$ at a given location during 10 s. $\langle \eta \rangle \simeq 0$. Forcing amplitude $U_{rms}=0.4$ V. Forcing frequency band $0\leq f \leq 6$ Hz.}
\label{fig01}
\end{figure}

A typical recording of the surface wave amplitude, $\eta(t)$, at a given location is displayed in the inset of Fig.\ \ref{fig01} as a function of time. The surface strongly fluctuates with a large distribution of amplitudes (see afterwards). The mean value of the amplitude is close to zero. In order to characterize the statistical properties of such a signal  (inset of Fig.\ \ref{fig01}), $\eta(t)$ is recorded by means of an acquisition card with a 1 kHz sampling rate during 3000 s, leading to $3 \times10^6$ points recorded. The power spectrum and the probability density function (PDF) are then computed. 

\begin{figure}[h]
\centerline{
\epsfysize=65mm
\epsffile{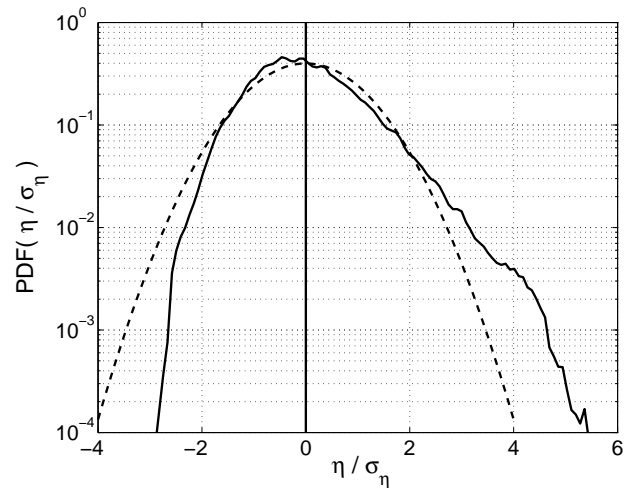} 
}
\caption{Probability density function of the normalized wave height, $\eta(t)/\sigma_{\eta}$. Standard deviation $\sigma_{\eta}\equiv \sqrt{\langle \eta^2\rangle}=2.6$ mm, flatness $\langle \eta^4\rangle/\langle \eta^2\rangle^{2}=4$, and skewness $\langle \eta^3\rangle/\langle \eta^2\rangle^{3/2}=0.65$. Gaussian fit with zero mean and unit standard deviation ($--$). Same forcing parameters as in Fig.\ \ref{fig01}. }
\label{fig02}
\end{figure}

At high enough forcing, the signature of a wave turbulence regime is observed \cite{Falcon06}: a scale invariant spectrum with two power-law frequency dependences (see Fig.\ \ref{fig01}) and an asymmetric PDF (see Fig.\ \ref{fig02}). The low frequency spectrum part $\sim f^{-4.3}$ corresponds to the gravity regime, whereas the high frequency one $\sim  f^{-2.8}$  corresponds to the capillary regime. For the present characteristics of the forcing, both power-law exponents are in fair agreement with weak turbulence theory predicting a power spectrum of the wave amplitudes $\sim f^{-4}$ for gravity waves \cite{Zakharov67Grav}, and $\sim  f^{-17/6}$ for capillary waves \cite{Zakharov67Cap}. However, the $f^{-4}$ scaling has been also ascribed to cusps \cite{Kuznetsov04}. In addition, as emphasized in \cite{Falcon06}, only the capillary regime is robust, the exponent for the gravity regime being strongly dependent of the characteristics of the forcing. The cross-over near 30 Hz corresponds to the transition between gravity and capillary wave turbulence spectra. At still higher frequencies (greater than 150 Hz), viscous dissipation dominates and ends the cascade of energy injected from large scale forcing.

The statistical distribution of wave height, $\eta$, at a given location is displayed in Fig.\ \ref{fig02}. At high enough forcing, the PDF is no longer Gaussian, and becomes asymmetric. The positive rare events such as high crest waves are more probable than deep trough waves. This also can be directly observed on the temporal signal $\eta(t)$ in Fig.\ \ref{fig01}. 

To test the intermittent properties of a stochastic stationary signal $\eta(t)$, one generally computes the increments $\delta\eta(\tau)\equiv \eta(t+\tau)-\eta(t)$. The structure functions of the signal $s_p(\tau) \equiv  \langle |\delta\eta(\tau)|^p \rangle = \langle|{\eta}(t+\tau)-{\eta}(t)|^p\rangle$ are also computed to seek a possible scaling behavior with the time lag $\tau$ \cite{Frisch95}, $\langle \cdot \rangle$ denoting a temporal average.  However, if a signal has a steep power spectrum $E(f) \sim f^{-n}$ with $n>3$ (e.g. in Fig.\ \ref{fig01}), the signal is then at least one time differentiable, and the increments are thus poorly informative (since they are dominated by the differentiable component), and $s_2(\tau)\equiv \langle[{\eta}(t+\tau)-{\eta}(t)]^2\rangle \sim \tau^2$ whatever $n$ \cite{Babiano85,Pope00}.  To test intermittency properties of such a signal, a more pertinent statistical estimator is related to the second-order differences of the signal, ${\Delta}\eta(\tau)\equiv {\eta}(t+\tau)-2{\eta}(t)+{\eta}(t-\tau)$ \cite{Workinprogress}. The structure functions are then defined as ${\mathcal S}_p(\tau) \equiv  \langle |\Delta\eta(\tau)|^p \rangle$. Note that other more complex estimators exist based on wavelet analysis \cite{Muzy93} or on inverse statistics \cite{Jensen99} of smooth signals. 

\begin{figure}[h]
\centerline{
\epsfysize=65mm
\epsffile{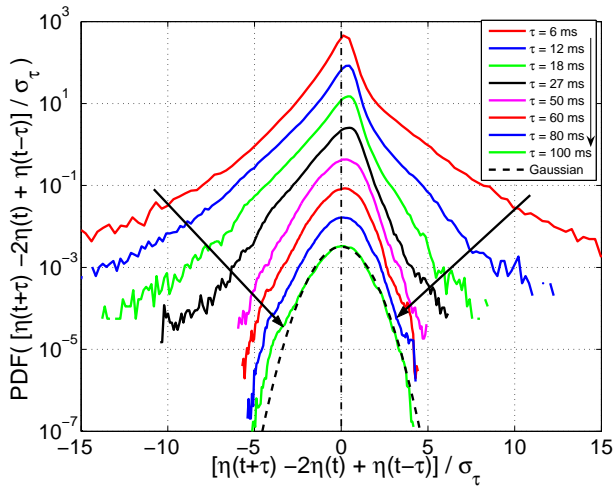} 
}
\caption{Probability density functions of second-order differences of the wave height $[\eta(t+\tau) -2\eta(t) + \eta(t-\tau)] / \sigma_{\tau}$ for different time lags $6 \leq \tau \leq 100$ ms (from top to bottom). Gaussian fit with zero mean and unit standard deviation ($--$). Correlation time $\tau_c\simeq 63$ ms. Each curve has been shifted for clarity. Same forcing parameters as in Fig.\ \ref{fig01}.}
\label{fig03}
\end{figure}

 The probability density functions of the second-order differences of the surface wave height, $\Delta\eta(\tau)$, normalized to their respective standard deviation $\sigma_{\tau}$, are plotted in Fig.\ \ref{fig03} for different time lags $6 \leq \tau \leq 100$ ms, the correlation time of $\eta(t)$ being $\tau_c\simeq 63$ ms. A shape deformation of the PDFs of $\Delta\eta(\tau)/\sigma_{\tau}$ is observed with the time lag $\tau$. The PDF is nearly Gaussian at large $\tau$. When $\tau$ is decreased from this integral scale, the PDF's shape changes continuously, and strongly differs from a Gaussian (see the PDF's tails in Fig.\  \ref{fig03}). This is a direct signature of intermittency. The extreme fluctuating events (large values of $\Delta\eta(\tau)/\sigma_{\tau}$) are all the more likely when the time scale $\tau$ is short. Thus, the signal of the surface-wave amplitude displays intermittent bursts during which the slope varies in an abrupt way within a short time. The second-order differences of the wave-amplitude signal are indeed related to intermittency of the local slope increments of the surface waves.
 
\begin{figure}[h]
\centerline{
\epsfysize=65mm
\epsffile{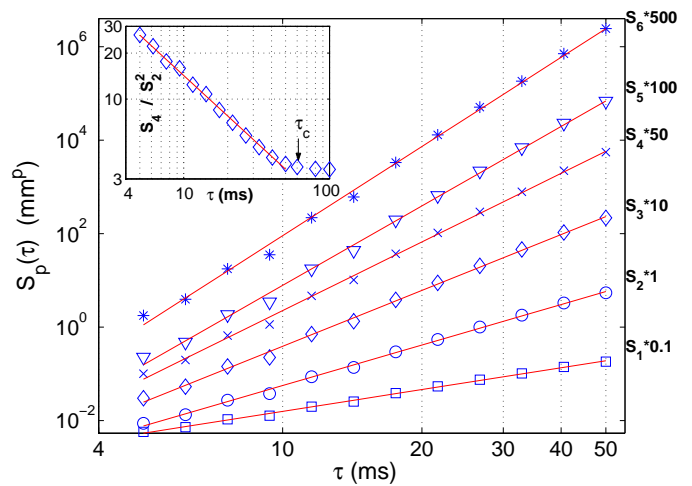}  
}
\caption{Structure functions ${\mathcal S}_p(\tau)$ of the second-order differences of the wave amplitude as functions of the time lag $\tau$, for $1\leq p \leq 6$ (as labeled).  ($-$): Power law fits, ${\mathcal S}_p(\tau) \sim \tau^{\xi_p}$, where the slopes $\xi_p$ depend on the order $p$ (see Fig.\ \ref{fig05}). Curves has been shifted for clarity. Inset: Flatness ${\mathcal S}_4/{\mathcal S}_2^2$ as a function of $\tau$.  ($-$): Power law fit with a slope $-0.88$. Correlation time $\tau_c \simeq 63$ ms. Same forcing parameters as in Fig.\ \ref{fig01}.}
\label{fig04}
\end{figure}

Figure\ \ref{fig04} shows the structure functions ${\mathcal S}_p(\tau)$ of the second-order differences of the wave amplitude as a function of the time lag $\tau$. For $5 \leq \tau \leq 50$ ms, all the structure functions of order $p=1$ to 6 are found to be power laws of $\tau$, ${\mathcal S}_p(\tau) \sim \tau^{\xi_p}$, where $\xi_p$ is an increasing function of the order $p$. When $\tau \gtrsim \tau_c$, ${\mathcal S}_p(\tau)$ is found to saturate (e.g. towards $2\langle \eta^2 \rangle$, for p=2) as usual \cite{Frisch95} (data not shown).  To quantify the intermittency of the signal (i.e. the PDF shape deformation across the temporal scales), the dependence of the flatness, ${\mathcal S}_4/{\mathcal S}_2^2$,  as a function of $\tau$ is displayed in the inset of the Fig.\ \ref{fig04}. At large $\tau$, the flatness is close to 3 (the value for a Gaussian) and increases up to 26 at the shortest $\tau$, corresponding to a much flatter PDF (see Fig.\ \ref{fig03}). The flatness is a power law of the time scale: ${\mathcal S}_4/{\mathcal S}_2^2 \sim \tau^{c}$ with $c=-0.88\pm0.03$. 

\begin{figure}[t]
\centerline{
\epsfysize=65mm
\epsffile{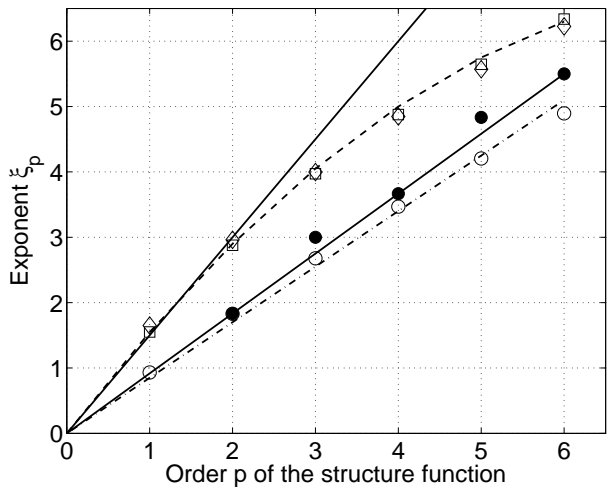} 
}
\caption{Exponents $\xi_p$ of the structure functions as a function of $p$. $\xi_p$ computed from the ($\square$) second-order differences (from the slopes of Fig.\ \ref{fig04}), or from the ($\Diamond$) third-order differences, and fitted by ($--$) $\xi_p=c_1p-\frac{c_2}{2}p^2$ with $c_1=1.65$ and $c_2=0.2$. ($-$) Theoretical prediction using dimensional analysis $\xi_p=3p/2$. ($\circ$) $\tilde{\xi}_p$ computed from the first-order increments, and fitted by ($-\cdot$) $\tilde{\xi}_p=0.85p$.  ($\bullet$) Theoretical points (from the first-order increments) \cite{Newell03} and dimensional estimate  $\tilde{\xi}_p=11p/12$ ($-$). }

\label{fig05}
\end{figure}

The evolution of the exponents $\xi_p$ of the structure functions as a function of $p$ is shown in Fig.\ \ref{fig05} from the slopes of the log-log curves in Fig.\ \ref{fig04}.  $\xi_p$ is found to be a nonlinear function of $p$ such that $\xi_p=c_1p-\frac{c_2}{2}p^2$ with $c_1=1.65\pm0.05$ and $c_2=0.2\pm0.02$. The value of $c_1$ is related to the exponent of the low frequency spectrum, $-2c_1 -1\approx -4.3 \pm 0.1$. As said above, this value of $c_1$ can be related to the cusps observed on the fluid surface. They correspond to discontinuities in the vertical velocity $v$ of the surface, thus leading to $f^{-2}$ spectrum, i.e., $\langle [v(t+\tau)-v(t)]^2\rangle \propto \tau$. This leads to the dimensional estimate $\langle |{\eta}(t+\tau)-2{\eta}(t)+{\eta}(t-\tau) |^p\rangle \propto  \tau^{p+p/2} = \tau^{3p/2}$, in fair agreement with the measurements for $p=1$ and $2$ (see Fig.\ \ref{fig05}). The nonlinearity of $\xi_p$ ($c_2\neq 0$) is another direct signature of intermittency \cite{Frisch95}. This intermittency is observed for $20 \leq 1/\tau \leq 200$ Hz, that is, for the capillary wave regime. The so-called intermittency coefficient $c_2$ can be also deduced from the measurement of the flatness as a function of $\tau$ (inset of Fig.\ \ref{fig04}). Indeed, inserting the expression of $\xi_p$ into $F={\mathcal S}_4/{\mathcal S}_2^2$ with ${\mathcal S}_p(\tau) \sim \tau^{\xi_p}$, leads to $F\sim \tau^{-4c_2}$. Thus, our measurements give $c_2=-c/4=0.22\pm0.008$ which is in agreement with the value of $c_2$ deduced from the exponents of the structure functions. Note that this intermittency coefficient is robust when using a third-order increment processing (see $\Diamond$-symbols in Fig.~\ref{fig05}).

Figure \ref{fig05} shows also the exponents $\tilde{\xi}_p$ of the structure functions computed from the first-order differences of the signal, $\langle |\eta(t+\tau) - \eta(t)|^p\rangle \sim \tau^{\tilde{\xi}_p}$, (open circles in Fig.\ \ref{fig05}). One can thus compare with theoretical predictions of weak turbulence \cite{Newell03} (solid circles in Fig.\ \ref{fig05}). Dimensional analysis for weak capillary wave  turbulence gives $\tilde{\xi}_p =11p/12$ (solid line).  Although the experimentally measured slope is slightly smaller than $11/12\simeq 0.92$, it is too close to 1 in order to display intermittency just by computing the structure functions from the signal increments. Indeed, as said above, since our signal spectrum is very steep, the signal increments are poorly informative, and the exponents $\tilde{\xi}_p$ of the structure functions of the signal increments, $\langle |\eta(t+\tau) - \eta(t)|^p\rangle \sim \tau^{\tilde{\xi}_p}$, are expected to be such that $\tilde{\xi}_p \sim p$ \cite{Babiano85,Pope00}. This shows that second-order differences should be computed in order to test intermittency properties.

We have reported that short-scale intermittency occurs on the second-order differences of surface wave amplitude. As previously proposed, intermittency could be related to coherent structures on the fluid surface \cite{Choi05}, such as wave breaking \cite{Yokoyama04} or whitecaps \cite{Connaughton03}. Here,  wave breaking or whitecaps do not occur, but cusps are observed on the fluid surface. However, we do not presently have a theory that determines $\xi_p$ for large $p$. We think that the observation of small scale intermittency in our system that strongly differs from high Reynolds number hydrodynamic turbulence is of primary interest. It can indeed motivate explanations of intermittency different than the ones considering the dynamics of the Navier-Stokes equation or the existence of coherent structures. A more general explanation can be related to the properties of the fluctuations of the energy flux that are shared by different systems displaying an energy cascade.

\begin{acknowledgments}
We gratefully aknowledge A. C. Newell for a series of lectures on wave turbulence given at ENS. We thank S. Roux, B. Audit and A. C. Newell for fruitful discussions at the early stage of this work. We thank N. Mordant for helpful comments. E. Falcon gratefully acknowledges the hospitality of the LPS at the ENS physics department.
\end{acknowledgments}


\begin{thebibliography}
\bibitem
\bibitem{Batchelor49}G. K. Batchelor and A. A. Townsend, Proc. R. Soc. Lond. A {\bf 199}, 238 (1949)
\bibitem{K62}A. N. Kolmogorov, J. Fluid Mech. {\bf 13}, 82 (1962)
\bibitem{Frisch95}U. Frisch, {\it Turbulence}, (Cambridge University Press, Cambridge, 1995) and references therein
\bibitem{Li05}Y. Li and C. Meneveau, Phys. Rev. Lett. {\bf 95}, 164502 (2005); L. Chevillard et al., Phys. Rev. Lett. {\bf 95}, 064501 (2005); L. Chevillard and C. Meneveau, Phys. Rev. Lett. {\bf 97}, 174501 (2006)
\bibitem{Falkovitch01}G. Falkovitch, K. Gawedzki, and M. Vergassola, Rev. Mod. Phys. {\bf 73}, 913 (2001)
\bibitem{Zakharov67Grav}V. E. Zakharov  and N. N. Filonenko, Phys. Dokl. {\bf 11}, 881 (1967); V. E. Zakharov  and M. M. Zaslavsky, Izv. Atm. Ocean. Phys. {\bf 18}, 747 (1982)
\bibitem{Zakharov67Cap}V. E. Zakharov and N. N. Filonenko, J. App. Mech. Tech. Phys. {\bf 8}, 37 (1967)
 \bibitem{Connaughton03}C. Connaughton, S. Nazarenko and A. C. Newell, Physica D {\bf 184}, 86 (2003)
\bibitem{Biven01} L. Biven, S. Nazarenko and A. C. Newell, Phys. Lett. A, {\bf 280}, 28 (2001); A. C. Newell, S. Nazarenko and L. Biven, Physica D {\bf 152-153}, 520 (2001); Y. V. Lvov and S. Nazarenko Phys. Rev. E {\bf 69}, 066608 (2004)
\bibitem{Choi05}Y. Choi, Y. V. Lvov, S. Nazarenko and B. Pokorni, Phys. Lett. A, {\bf 339}, 361 (2005)
\bibitem{Yokoyama04}N. Yokoyama, J. Fluid Mech. {\bf 501}, 169 (2004) 
\bibitem{Toba73}Y. Toba, J. Oceanog. Soc. Japan {\bf 29}, 209 (1973); K. K. Kahma, J. Phys. Oceanogr. {\bf 11}, 1503 (1981); G. Z. Forristall, J. Geophys. Res. {\bf 86}, 8075 (1981); M. A. Donelan et al., Phil. Trans. R. Soc. Lond. A {\bf 315}, 509 (1985)
\bibitem{Wright96}W. B. Wright, R. Budakian and S. J. Putterman, Phys. Rev. Lett. {\bf 76}, 4528 (1996); W. B. Wright, R. Budakian, D. J. Pine and S. J. Putterman, Science {\bf 278}, 1609 (1997)
\bibitem{Lommer02}M. Lommer and M. T. Levinsen J. Fluoresc. {\bf 12}, 45 (2002); E. Henry, P. Alstr{\o}m and M. T. Levinsen, Europhys. Lett. {\bf 52}, 27 (2000)
\bibitem{Brazhnikov02}M. Yu. Brazhinikov, G. V. Kolmakov and A. A. Levchenko, Sov. Phys JETP {\bf 95}, 447 (2002); M. Yu. Brazhinikov et al., Europhys. Lett. {\bf 58}, 510 (2002); G. V. Kolmakov et al. Phys. Rev. Lett. {\bf 93}, 074501 (2004)
\bibitem{Onorato04}M. Onorato et al. Phys. Rev. E {\bf 70}, 067302 (2004)
\bibitem{Falcon06}E. Falcon, C. Laroche, S. Fauve Phys. Rev. Lett (in press)
\bibitem{Kuznetsov04} E. A. Kuznetsov, JETP Letters {\bf 80}, 83 (2004)
\bibitem{Babiano85}A. Babiano, C. Basdevant and R. Sadourny, J. Atmos. Sci. {\bf 42}, 941 (1985); 
\bibitem{Pope00}S. B. Pope, {\it Turbulent Flows}, (Cambridge University Press, 2000); A. S. Monin and A. M. Yaglom, {\it Statistical Fluid Mechanics: Mechanics of Turbulence, Vol.2}, (The MIT Press, 1975); P. A. Davidson and B. R. Pearson, Phys. Rev. Lett., {\bf 95}, 214501 (2005)
\bibitem{Workinprogress}E. Falcon, S. Roux and B. Audit, in preparation; L. Biferale, M. Cencini, A. S. Lanotte and D. Vergni, Phys. Fluids {\bf 15} 1012 (2003)
\bibitem{Muzy93}J.-F. Muzy, E. Bacry, and A. Arneodo Phys. Rev. E {\bf 47}, 875 (1993); Phys. Rev. Lett. {\bf 67}, 3515 (1991)
\bibitem{Jensen99}M. H. Jensen, Phys. Rev. Lett. {\bf 83}, 76 (1999); L. Biferale et al., Phys. Rev. Lett. {\bf 87}, 124501 (2001)
\bibitem{Newell03}L. J. Biven, C. Connaughton and A. C. Newell, Physica D {\bf 184}, 98 (2003)




\end{thebibliography}
\end{document}